\documentstyle[12pt]{article}
\begin{document}

\begin{center}
PHYSICS OF HEAVY QUARKS\\
\vspace*{1cm}
Kacper Zalewski\footnote{Also at the Institute of Nuclear Physics, Cracow.\\
This work was partly supported by the KBN grant 2P 302 076 07}\\
Institute of Physics, Jagellonian University,\\ ul. Reymonta 4,
 30 059 Krak\'ow, Poland\\
\end{center}

\vspace*{2cm}

Selected problems in heavy quark physics are discussed. The wealth of research
problems in this field of physics is stressed.

\section{Introduction}

Heavy quark physics is a broad and active field of particle physics. Within it,
hundreds of theoretical papers are produced every year and the production rate
keeps increasing. In this short presentation I shall concentrate on recently
obtained insights and on open problems. The experimental data quoted without
giving the source are either from the 1994 Tables of the Particle Data Group
\cite{PDG}, or from the EPS Conference held in Brussels this summer.

According to the standard model there are six kinds of quarks. In order of
increasing mass they are denoted $u,\;d,\;s,\;c,\;b,\;t$. The last three are
considered heavy, because their masses are much larger than $\Lambda_{QCD}$,
i.e. than about $0.5$ GeV. The mass of the $c$-quark can be roughly estimated
as half the mass of the lightest $c\overline{c}$ quarkonium, which gives $m_c
\approx 1.5$ GeV. This in fact it not very heavy --- only about three times
$\Lambda_{QCD}$. The mass of the recently discovered $t$ quark is $m_t = (180
\pm 12)$ GeV, which implies that the $t$-quark decays, usually into a $W$-boson
and a $b$-quark, before it has time to hadronize. Consequently, the physics of
the $t$-quarks is already well understood. The mass spectrum of the heavy
quarks causes that most of the new ideas apply best to $b$-quarks. For
$t$-quarks, the problems are fewer and they can be usually solved without
making
controversial assumptions. For $c$ quarks, we are too far from the heavy quark
limit, where all the quantities of order $\Lambda_{QCD}$ can be neglected
compared to the mass of the heavy quark. On the long run this may make the
physics of the $c$-quarks more interesting than the physics of the $b$-quarks,
but for the moment it is often just too difficult.

Let us begin by considering the problem: what is meant by the quark mass?

\section{Quark masses}

The standard definition of mass, $m = \sqrt{E^2 - \vec{p}^2}$ is not applicable
to quarks, because the energy $E$ and the momentum $\vec{p}$ on its
left-hand-side should be measured for free particles. Looking for a free quark
not interacting with other couloured objects is like looking for one end of a
string not attached to another end. There is no chance for success. The next
choice, when trying to define the quark mass, is to make use of the mass
parameter $m_0$ from the Lagrangian. This, however, has corrections. The fact
that the quark is part of the time a quark-gluon system (the contribution of
the gluon loop on the quark line) changes the mass by

\begin{equation}
\Sigma^{(1)} = \frac{m \alpha_s(\mu)}{\pi}\left[ \frac{1}{\varepsilon} - \gamma
+ \log(4\pi) + \log\frac{\mu^2}{m^2} + \frac{4}{3} \right],
\end{equation}
where $\gamma = 0.5772\ldots$ is Euler's constant. From this formula one sees
two difficulties; moreover, there are two others not directly visible.
\begin{itemize}
\item The limit $\varepsilon \rightarrow +0$ should be taken, thus the formula
as it stands does not make sense.
\item The scale parameter $\mu$ is arbitrary.
\item The formula has been obtained using dimensional regularization. There are
many other methods of regularizing (various cut off procedures, putting the
theory on a lattice etc.), which yield different formulae.
\item This correction is only the first term of an infinite series, in general
convergence problems are expected.
\end{itemize}.

The infinity is eliminated by replacing the mass $m_0 + \Sigma^{(1)}$ by the
obviously equal number $(m_0 + \delta m) + (\Sigma^{(1)} - \delta m)$. The
trick is to choose $\delta m$ so that it cancels the infinity in
$\Sigma^{(1)}$. Since nothing is known about $m_0$, one can assume that
$\Sigma^{(1)}$ does not introduce an infinity in the first term. This recipe
leaves much freedom in the choice of $\delta m$. Choosing $\delta m =
\frac{1}{\varepsilon}$ one gets the so called minimal subtraction mass.
Including in $\delta m$ also $-\gamma + \log(4\pi)$, which is convenient, one
obtains the very popular $\overline{m}$ mass known as the MS-bar mass. Choosing
$\delta m = \Sigma^{(1)}$ one obtains the pole mass $m^{P}$ etc. Each of these
masses depends on the scale $\mu$. This arbitrary scale is usually chosen of
the order of the mass of the quark being considered. For instance, the Particle
Data Group \cite{PDG} tabulates the quark masses $\overline{m}(\overline{m})$.
The differences between the various masses are significant. For instance, using
the formula for $\Sigma^{(1)}$ one finds for quark $Q$

\begin{equation}
\overline{m}_Q(\overline{m}_Q) = m^P_Q\left(1 - \frac{4
\alpha_s(\overline{m}_Q)}{3 \pi} \right).
\end{equation}
Typical values of $\alpha_s(m_Q)$ for the heavy quarks are $0.35,\;0.20,\;0.10$
for the $c,\;b,\;t$ quarks respectively. This gives in the present (very crude)
approximation the differences between the pole masses and the MS-bar masses
$0.17$ GeV, $0.34$ GeV and $7$ GeV. More careful calculations give for the $c$
and $b$ quarks $0.26$ GeV and $0.51$ GeV \cite{TYN}, while typical values for
the $t$-quark are $(8 \mbox{---} 9 )$ GeV. An obvious question is: what is the
mass found in Fermilab for the $t$ quark? The description of the measurement
provides an unambiguous operational definition of this mass, but to which
of the theoretical mass parameters does it correspond? Somewhat surprisingly
this problem is still controversial. The pole mass, however, seems to be the
most popular interpretation.

For the other normalization schemes it is possible to perform analogous
analyses, therefore, the existence of various renormalization schemes is not a
serious difficulty.

The convergence problem, however, has been recently found to introduce an
interesting complication. References can be traced starting from the recent
review \cite{SAC}. One finds (if one uses dimensional regularization) that the
series used to define the pole mass is divergent. It can be used as an
asymptotic series, but then it defines the pole mass only approximately, with
an error of about $50$ MeV. This is the reason why the MS-bar masses are now
the popular ones for the $c$ and $b$ quarks. For the $t$ quark the situation is
different. With present experimental uncertainties an additional uncertainty of
$50$ MeV is irrelevant. On the other hand, the relation between the pole mass
and the MS-bar mass has a much greater uncertainty. The calculations necessary
to reduce this uncertainty are possible, but so hard that they have not yet
been done and are unlikely to be performed it the nearest future. Therefore, if
the measured mass is the pole mass, expressing it in terms of the MS-bar mass
would be an unnecessary loss of precision.

\section{Heavy particles}

By heavy particles we mean here particles containing one or more heavy quarks
or antiquarks. The best studied case is the nonrelativistic approximation for
the quarkonia $\overline{Q}Q$. In particular for bottomonia, it is possible to
get a very good fit to the masses (averages only for the $P$-states) below the
threshold for strong decays, for the leptonic widths and for the dipole
transition matrix elements. One can use the nonrelativistic Schr\"odinger
equation with the simple spherically symmetrical potential

\begin{equation}
V(r) = a\sqrt{r} + \frac{b}{r} + c,
\end{equation}
where $a,\;b,\;c$ are constants \cite{MZA}. How to make a relativistic theory
is still controversial.

For heavy particles containing light quarks the situation is more difficult,
because for them the nonrelativistic theory does not make much sense. A break
through has been the idea to use expansions in the inverse of the heavy quark
mass. For instance, for the mass of a particle with one heavy quark $Q$ one
finds

\begin{equation}
M_H = m_Q + \overline{\Lambda} + \frac{\langle\vec{p}^2\rangle}{2 m_Q} +
\frac{\langle\vec{\sigma}\cdot\vec{B}\rangle}{2 m_Q} +
 \frac{1}{m_Q^2}\left[ \mbox{Darwin + Spin-orbit + IterII} \right].
\end{equation}

The leading term is just the mass of the heavy quark. The term of order
$m_Q^0$, denoted $\overline\Lambda$, is the energy of the light component in
the colour-field of the heavy quark. The heavy quark is here considered as a
static source of potential. Note the generality of this formulation. The light
component may be an antiquark, as in valence models of $Q\overline{q}$
mesons, a pair of quarks, as in the valence models of $Qqq$ barions, or a more
complicated combination of light quarks, light antiquarks and gluons, as in
some more sophisticated models. The corrections of order $O(m_Q^{-1})$
correspond to the kinetic energy of the heavy quark and to the Pauli
interaction of the magnetic moment of the heavy quark with the chromomagnetic
field created by the light component. The magnetic term is responsible for the
hyperfine mass splittings in the mass spectra. For instance the difference
between the $B$ meson and the $B^*$ meson is that in the first the spins of the
heavy meson and of the light component give the resultant spin of the particle
equal zero, while in the second this spin equals one. One finds

\begin{equation}
\langle\vec{\sigma}\cdot\vec{B}\rangle = \frac{3}{4}\left(M^2_{B^*} -
M^2_B\right) \approx 0.37 GeV^2..
\end{equation}
Since this average should not depend on the mass of the heavy quark, one
expects a similar value for the $(D,\;D^*)$ system. In fact the experimental
number is $0.41$ GeV$^2$. This can be formulated differently: the experimental
fact that the hyperfine splitting for $Q=b$ is about three times smaller than
the hyperfine splitting for $Q=c$, is explained here as a consequence of the
fact that the $c$-quark is about three times lighter than the $b$-quark. The
kinetic energy term has no such direct connection to experimental data and,
therefore, its value is controversial. It can be shown that $\langle \vec{p}^2
\rangle > \langle\vec{\sigma}\cdot\vec{B}\rangle$ (\cite{BIG} and references
given there) and typical estimates are between this lower limit and its double.
For the terms of order $m_Q^{-2}$ we have given only the names. The first two,
the Darwin term and the spin-orbit interaction, are familiar from the Dirac
theory of the hydrogen atom. The third term is the second perturbative
iteration of the $O(m_Q^{-1})$ term.

One can apply this approach also to higher resonances. When the light component
consists of a light antiquark in a $P$ state, its angular momentum can be
$\frac{1}{2}$ or $\frac{3}{2}$. The parity is plus. Combining that with the
spin $\frac{1}{2}$ and positive parity of the heavy quark, one finds four
excited states with spins and parities: $0^+,\;1^+,\;1^+,\;2^+$. Experimentally
one finds two charmed mesons $D_1^{**}$ and $D_2^{**}$ with masses $(2423 \pm
3)$
MeV and $(2458 \pm 2)$ MeV respectively and one bottom meson $B^{**}$ with mass
$(5733 \pm 17)$ MeV. A $D^{**}$ meson decays into a pion and a $D$ or $D^*$
meson. Using angular momentum and parity conservation, as well as the
information that the pion is produced from the light component, one can see
that the mesons with the angular momentum of the light component equal
$\frac{1}{2}$, decay producing a pion in an $S$-state. Such mesons are broad
and difficult to observe. The $D^{**}$ mesons with the angular momentum of the
light component equal $\frac{3}{2}$, on the other hand, produce pions in
$D$-states and are narrow, because of the suppression of the decay probability
by the angular momentum barrier. This explains, why only two $D^{**}$ mesons
have been observed. The hyperfine splitting between these mesons is about $30$
MeV. Since this is an effect of order $O(m_Q^{-1})$, the corresponding
splitting for the $B^{**}$ mesons is expected to be about $10$ MeV, and indeed
cannot be seen at the present resolution of $17$ MeV. This explains, why for
the moment only one $B^{**}$ meson has been seen. One also can predict that in
order to distinguish the two $B^{**}$ mesons, the resolution will have to be
improved by about a factor of two.

\section{Decays of heavy particles}

Decays of heavy particles are an important source of information about the
elements of the Cabibbo-Kobayashi-Masakawa (CKM) matrix. From the point of view
of the standard model these matrix elements are coupling constants (not all
independent from each other!) as fundamental as e.g. the electron charge. Where
these constants are known, comparison of the theoretical predictions with
experiment yields interesting tests of the standard model.

Let us consider the semileptonic decay $\overline{B} \rightarrow D^* e
\overline{\nu}$. In this decay the $b$-quark, with a probability amplitude
proportional to the $CKM$ matrix element $V_{cb}$, goes over into a $c$-quark.
In the process it emits a virtual $W^-$ intermediate boson, which decays into
the $e^-,\;\overline\nu$ pair. The problem is to extract the modulus $|V_{cb}|$
from the experimental data.

In the heavy quark limit the heavy mesons $\overline{B}$ and $D^*$ are similar
to hydrogen atoms. In each case the heavy quark sits in the middle, like the
proton in hydrogen, and the light component surrounds it, like the
electron cloud surrounds the proton in the hydrogen atom. The energy and
momentum of the $W$-boson are very large on the scale of the momenta of the
light components. An analogy would be a $1$ MeV photon hitting the proton in
hydrogen. In this situation the heavy "nucleus" behaves as if it were free. It
gets ejected with large momentum (on the scale of the light stuff) from its
original position. The $b$-quark absorbing (or equivalently
emitting) the $W$-boson changes into a $c$-quark. Note that since the $c$-quark
 is very heavy, large momentum does not necessarily mean large velocity.
This process, however, is not yet the process $\overline{B} \rightarrow D^*$.
In order to get the probability amplitude for this decay it is necessary to
multiply the probability amplitude for the ejection of the heavy quark by the
probability amplitude that the light component of the original
$\overline{B}$-meson will reorganize itself into the light component of the
recoiling $D^*$-meson. This is given by the overlap of the two corresponding
wave functions. Thus, omitting the less interesting (known) terms, the decay
amplitude is

\begin{equation}
A = V_{cb} \overline{u}_{\vec{v}'}\gamma_\mu(1-\gamma^5)u_{\vec{v}}F(\omega).
\end{equation}
Here $\vec{v}$ and $\vec{v}'$ denote the initial and the final velocities of
the heavy quark. In the heavy quark limit these velocities are equal to the
velocities of the corresponding mesons. The argument $\omega = v^\mu v'_\mu$,
which can be interpreted as the Lorentz factor of the $D^*$ as seen in the rest
frame of the $\overline{B}$, is a measure of the recoil velocity. The overlap
factor, known as the Isgur-Wise function, is

\begin{equation}
F(\omega) = \int \psi^*_{v'}(\vec{r})\psi_v(\vec{r}) d^3r.
\end{equation}
Note that the overlapping wave functions of the light components differ only by
the velocities of their centres. The change of the $b$-quark into a $c$-quark
and the change of the relative spin orientation of the heavy and light quarks
from antiparallel to parallel have in leading order no effect on the wave
function of the light component. The remaining difficulty is how to extract
from the data the factor $|V_{cb}|$ without using a specific model for the
Isgur-Wise function.

Two solutions to this problem have been proposed. In the exclusive approach one
notices that for $\vec{v} = \vec{v}'$ the two overlapping functions are
identical and that consequently $F(1) = 1$ from the normalization of the wave
function. In this approach one obtains from the data the product
$|V_{cb}|F(\omega)$ and extrapolates it to zero recoil, where $F(\omega = 1) =
1$. In the inclusive approach, one gives up the constraint that the final
charmed state must be a $D^*$ meson. Then the Isgur-Wise function is replaced
by the probability amplitude that the light component will reorganize itself
into anything, which is, of course, equal one. Thus, one uses data for the
inclusive process $\overline{B} \rightarrow X_c e\overline\nu$. Here $X_c$
denotes any state containing the quark $c$. Since the $b$-quarks decay almost
always into $c$-quarks, $X_c$ can in practice be replaced by $X$ meaning
anything. We have presented here only the leading term analysis. In practice
one includes various corrections, which are still somewhat controversial.
Fortunately they change the calculated values of $|V_{cb}|$ by only a few
percent. Incidentally, the analogous problem of extracting the $CKM$ matrix
element $|V_{ub}|$ from the data is much harder and is an active subject of
research.

Let us mention two open problems connected with inclusive decays (cf. e.g.
\cite{BIG2}). Theoretically one finds that the life times of the heavy
particles containing single $b$-quarks are well estimated using the spectator
model, i.e. neglecting the effect of the light components on the life times.
This corresponds to equal life times for all such particles. It is possible to
calculate corrections to this result and they turn out to be of a few percent.
This agrees well with experiment for meson decays, but for $\Lambda_b$ the
experimental life time is only $(0.72 \pm 0.06)$ of the $b$-quark life time
inferred from meson decays. The theoretical expectation for this ratio is below
one, but almost surely above 0.9. The second problem is the measured fraction
of the $B$ mesons, which decay semileptonicaly. Theory can reproduce it, but
at the condition that a large fraction of these decays leads to $\overline{c}c$
pairs. The average number of $c$ and $\overline{c}$ quarks per decay is
experimentally $(1.13 \pm 0.05)$, while the theoretical number necessary to get
agreement with the semileptonic branching ratio is $1.3$. This difference may
seem small, but it should be kept in mind that one $c$-quark is present in
almost every $b$-decay. Thus what counts is the surplus over this number. Here
the experimental number is less than half the theoretical one.

Finally let us mention the so called rare decays, i.e. the decays, where the
$b$-quark goes over into an $s$-quark and a photon, or lepton pair. Here the
theory involves pingwin diagrams, is quite complicated and is still being
refined, but what is important is that it agrees well with experiment. This
eliminates many ideas concerning "new physics" i.e. physics beyond the standard
model.

\section{Production of heavy particles}

 Heavy particle production is a broad and active subject. Here we shall only
mention a few problems, which now are attracting particular interest.

The calculated cross-section for the process $p\overline{p} \rightarrow
t\overline{t} X$ at the Tevatron is somewhat lower than measured. Since the
experimental uncertainties are large, however, and since the discrepancy
decreases as data improve, this does not seem to be a serious problem.

The ratio of the decay probability of $Z^0$ into $b\overline{b}$ to the decay
probability of $Z^0$ into any hadrons should be about 0.2, because there are
five kinds of quarks into which a $Z^0$ can decay and they all have masses
negligible compared to the $Z^0$ mass. Experimentally

\begin{equation}
R_b = \frac{\Gamma(Z^0 \rightarrow b\overline{b})}{\Gamma(Z^0 \rightarrow
\mbox{hadrons})} = 0.2205 \pm 0.0016
\end{equation}
in agreement with this crude estimate. Precise calculations, however, give $R_b
= 0.2155$, i.e. a ratio smaller by about three standard deviations than the
experimental one. This is considered as a possible problem for the standard
model. It is interesting that supersymmetry can increase the predicted $R_b$ so
that it becomes lower than the experimental value by only about one standard
deviation. If this is the correct explanation of the discrepancy, the lightest
supersymmetric particles should have masses below $100$ GeV and there is a good
chance of discovering them in the upgraded LEP accelerator. This is, of course,
a bold speculation, but it has recently triggered much discussion.
Incidentally, the corresponding ratio $R_c = 0.154 \pm 0.07$, to be compared
with the theoretical prediction $0.172$. Here, however, the experiment is very
difficult and a modification of the theory is not plausible, therefore this
discrepancy is expected to disappear, when data improves.

Finally let us mention the production of charmonia at the Tevatron. According
to theory those charmonia, which are not decay products of particles containing
$b$-quarks, should be mostly produced in gluon-gluon interactions. Such
interactions are much more likely to produce $P$-wave charmonia ($\chi$-states)
than $S$-wave charmonia ($\psi$-states). Therefore, the prediction was that the
direct production of $\psi$-states will be small and that a large majority of
such charmonia will come from decays of $\chi$-states. Experimentally it seems
that the direct production of $\psi$-charmonia is much stronger than expected,
sometimes stronger by more than an order of magnitude. One way out of this
difficulty is to assume that the $c\overline{c}$ systems in octet colour states
are an important intermediate state.


\begin{thebibliography}{999}

\bibitem{PDG}Particle Data Group, {\it Phys. Rev.} {\bf D50} (1994) 1173.

\bibitem{TYN}S. Titard and F.J. Yndurain, {\it Phys. Rev.} {\bf D49} (1994)
6007.

\bibitem{SAC}C.T. Sachrajda, {\it Acta Phys. Pol.} {\bf 26B} (1995) 731.

\bibitem{MZA}L. Motyka and K. Zalewski, {\it Acta Phys. Pol.} {\bf 26B} (1995)
829; {\it Zs. f. Phys.} {\bf C} in print.

\bibitem{BIG}I.I. Bigi et al., CERN preprint CERN-TH 7250/94 (1994).

\bibitem{BIG2}I.I. Bigi, {\it Acta Phys. Pol.} {\bf 26B} (1995) 641.

\end{thebibliography}
\end{document}